\documentclass[aps,prb,twocolumn,groupedaddress,showpacs]{revtex4}

\usepackage{graphics}
\usepackage{amssymb}

\begin{document}

\title{Stability of K-montmorillonite hydrates: Hybrid MC simulations}

\author{G. Odriozola}
\email[]{godriozo@imp.mx} \email[]{Tel: +5255 91758176}

\author{J. F. Aguilar}

\affiliation{Programa de Ingenier\'{\i}a Molecular, Instituto
Mexicano del Petr\'{o}leo, L\'{a}zaro C\'{a}rdenas 152, 07730
M\'{e}xico, D. F., M\'{e}xico}

\date{\today}
\begin{abstract}
$NP_{zz}T$ and $\mu P_{zz}T$ simulations of K-montmorillonite
hydrates were performed employing hybrid Monte Carlo simulations.
Two condition sets were studied, P=1 atm and T= 300 K (ground
level conditions), and P=600 atm and T= 394 K; this last condition mimics a
burial depth close to 4 km. For these conditions, swelling curves
as a function of the reservoir water vapor pressure were built. We
found the single layer K-montmorillonite hydrate stable for high
vapor pressures for both, burial and ground level conditions. A
simple explanation for this high stability is given.

\end{abstract}

\maketitle

\section{Introduction}
\label{intro}Clays are layer type aluminosilicate minerals,
existent everywhere in nature and industry, hence the importance
of a detailed understanding of their physics and chemistry. They
are used as building materials, ceramics, and catalysts; they are
employed in cosmetics, as rheological modifiers for paints, and in
technological processes such us oil well drilling. In this last
application, the control of their stability is a key for drilling
success. During it, the use of water based muds induce
destabilization of shale and clay formations that would
disintegrate, or heave, upon contact with water.

One way to maintain stability of shales during the drilling
process is by addition of potassium salts to drilling muds. This
helps to avoid fluid loss and water infiltration. These kind of
muds, that contain potassium ions dissolved in the water phase,
are widely used for drilling water-sensitive shales, specially
hard, brittle shales. Potassium cations in these systems replace
ions such as sodium found in most shales to produce less hydrated
clays with significantly reduced swelling potential. These ions
also help to hold the cuttings together, minimizing their
dispersion into finer particles. From all of the previous facts, a
good understanding of the role of potassium in the swelling clays
such as montmorillonite is mandatory, in particular at basin
conditions of hard experimental implementation.

Besides the experimental studies, computer simulations are
essential components of research on clay-water-cation systems
\cite{Boek95b,Skipper95c,Skipper97,Sposito01,Marry02,Young00,dePablo01,dePablo01b,Boek03,Hensen02,Odriozola_lan,Whitley04,Smith04}.
These simulations give microscopic insights that are difficult to
access experimentally. There are, however, not many computational
studies on the swelling of montmorillonite hydrates for potassium
interlayer cations
\cite{Boek95a,Skipper98,Chang99,Sposito99,Park00,Hensen01,Tambach04}.
In addition, excluding Hensen {\em et~al.}
\cite{Hensen01,Hensen02} and Tambach {\em et~al.} \cite{Tambach04}
works, these papers report solely $NP_{zz}T$ and $NVT$
simulations, where the number of water molecules per interlaminar
space is someway arbitrarily fixed, and hence, they do not show
the whole picture of swelling. A good example of this is that
hysteresis is naturally predicted by sampling in an open ensemble
\cite{Hensen02,dePablo01,Odriozola_jcp}, without the need of
measuring properties such as water chemical potential
\cite{Odriozola_lan}, immersion energies \cite{Young00}, or
swelling free energies \cite{Tambach04b,Whitley04}. Finally, this
paper focuses on the stability of the different hydrates in
contact with several reservoirs, which differ in temperature,
pressure, and water activity.

The importance of potassium as swelling inhibitor of clays and the
above mentioned reasons motivated us to study the microscopic
mechanisms underlying the behavior of K-montmorillonite hydrates
at equilibrium with different reservoirs. We performed simulations
of these systems in the $NP_{zz}T$ and $\mu P_{zz}T$ ensembles,
considering explicitly two clay layers in the simulation box to
avoid finite size effects. The simulations were carried out for
two condition sets. One at ground level, with P$=1$ atm and
T$=298$K, and the other one with P$=600$ atm and T$=394$K, which
corresponds to an average burial depth close to 4 km.

The paper is organized as follows. In Sec.~\ref{methods}, we
briefly describe the models and the methodology employed for
performing the simulations. The results are shown in
Sec.~\ref{results}. Finally, Sec.~\ref{summary} discusses the main
results and extracts some conclusions.

\section{Methodology}
\label{methods}
\subsection{The model}
A $4 \times 2$ layer of Wyoming type montmorillonite clay was
built up by replication of the unit cell given by Skipper {\em
et~al.} \cite{Skipper95a}. This layer has $L_x=21.12$
$\hbox{\AA}$, $L_y=18.28$ $\hbox{\AA}$ and $L_z=6.56$ $\hbox{\AA}$
dimensions. The Wyoming type montmorillonite was obtained by
isomorphous substitutions of trivalent Al atoms of the octahedral
sites by divalent Mg atoms, and tetravalent Si by trivalent Al
atoms. The unit cell formula of this clay is given by
K$_{0.75}$$n$H$_2$O(Si$_{7.75}$Al$_{0.25}$)(Al$_{3.5}$Mg$_{0.5}$)O$_{20}$(OH)$_4$
\cite{dePablo01}. Size effects were avoided by considering two
layers in the simulation box \cite{dePablo01}. Periodic boundary
conditions were imposed on the three space directions. The initial
configuration consists of water molecules randomly placed in the
interlaminar spaces, and six potassium ions distributed in the
interlayer midplanes. These counterions balance the negative
charge of the clay framework keeping the system electroneutral.

The rigid TIP4P model was used for water molecules
\cite{Jorgensen79,Tambach04} and the water clay interactions were
taken from Boek {\em et~al.}\cite{Boek95b} Here, site to site
intermolecular interactions are given by electrostatic and
Lennard-Jones contributions,
\begin{equation}\label{pot}
U_{ij}\!=\!\sum_{a,b}\! \left[ \frac{q_aq_b}{r_{ab}}+ 4
\epsilon_{ab} \left[ \left( \frac{ \sigma_{ab}}{r_{ab}}
\right)^{12} - \left( \frac{ \sigma_{ab}}{r_{ab}} \right)^{6}
\right] \right]
\end{equation}
where subindexes $i$ and $j$ are for molecules, and $a$ and $b$
run over all sites of each molecule. $q_{a}$ and $q_{b}$ are the
corresponding site charges, $\epsilon_{ab}$ and $\sigma_{ab}$ are
site to site specific Lennard-Jones parameters and $r_{ab}$ is the
inter-site distance. The Lennard-Jones parameters for single sites
are shown in Table \ref{parameters}. Here, those parameters for Si
were taken from Marry {\em et~al.} \cite{Levesque02}, and
parameters for Al and Mg were assumed to be equal to those of Si.
The site to site Lennard-Jones parameters are given by the
Lorentz-Berthelot rules
\begin{equation}
\sigma_{ab}=\frac{\sigma_a + \sigma_b}{2},
\end{equation}
\begin{equation}
\epsilon_{ab}=\sqrt{\epsilon_a \epsilon_b}
\end{equation}

\begin{table}
\caption{\label{parameters} Lennard-Jones parameters for
H$_2$O-clay-K interactions. Exceptions are the K-O and K-H
interactions.}
\begin{ruledtabular}
\begin{tabular}{ccc}
$\;\;\;\;$ Sites & $\epsilon$ (kcal/mol) & $\sigma$ ($\hbox{\AA}$) $\;\;\;\;$ \\
\hline $\;\;\;\;$ O & 0.1550 & 3.1536 $\;\;\;\;$ \\
$\;\;\;\;$ H & 0.0000 & 0.0000 $\;\;\;\;$ \\
$\;\;\;\;$ K & 3.630 & 2.4500 $\;\;\;\;$ \\
$\;\;\;\;$ Si & 3.150 & 1.840 $\;\;\;\;$ \\
$\;\;\;\;$ Al & 3.150 & 1.840 $\;\;\;\;$ \\
$\;\;\;\;$ Mg & 3.150 & 1.840 $\;\;\;\;$ \\
\end{tabular}
\end{ruledtabular}
\end{table}

On the other hand, the K-H$_2$O interactions and those between the
oxygens of the clay and potassium ions are based on the ones
proposed by Bounds \cite{Bounds85}. Bounds potential is chosen
since, while simple, it produces K-TIP4P radial distribution
functions in agreement with available experimental data and close
to those obtained by hybrid quantum mechanics/ molecular mechanics
(QM/MM) simulations, which naturally account for the many body
contributions to the potential \cite{Tongraar98}. That is, the K-O
radial distribution function peaks at 2.86 $\hbox{\AA}$ leading to
a first shell oxygen coordination number of 7.6, while high
accuracy QM/MM simulations performed at density functional theory
level (LANL2DZ basis set) give 2.81 $\hbox{\AA}$ of K-O distance
and 8.3 of coordination number. Experimental results give K-O
distances between 2.7 and 3.1 $\hbox{\AA}$ and coordination
numbers in the wide range of 4-8 \cite{Neilson85,Ohtaki93}. He
fitted the following pair potential for the K-H$_2$O
dispersion-repulsion contribution obtained from ab-initio
calculations,
\begin{eqnarray}
U_{K-H_2O} \!\!&=& \!\! A_{KO}\exp{(-b_{KO}r_{KO})}-C_{KO}/r_{KO}^4 \nonumber \\
& & \!\! -D_{KO}/r_{KO}^6+A_{KH}\exp{(-b_{KH}r_{KH_1})} \nonumber \\
& & \!\! +A_{KH}\exp{(-b_{KH}r_{KH_2})} \label{Bounds},
\end{eqnarray}
yielding $A_{KO}=$ 53884.0 kcal/mol, $b_{KO}=$ 3.3390
$\hbox{\AA}^{-1}$, $C_{KO}=$ 438.0 kcal$\hbox{\AA}^4$/mol,
$D_{KO}=$-638.0 kcal$\hbox{\AA}^6$/mol, $A_{KH}=$ 5747.0 kcal/mol
and $b_{KH}=$ 3.4128 $\hbox{\AA}^{-1}$. This intersite potential,
although somewhat more complicated than the Lennard-Jones type,
produces a much better match to the ab-initio data
\cite{Bounds85}. This is clearly seen if one tries to fit equation
(\ref{Bounds}) with a Lennard-Jones type potential. In fact,
parameters shown in table \ref{parameters} for potassium were
obtained by this way, yielding a not very good match although
reproducing the depth and position of the pair-potential minimum.
We observed that the discrepancies are very pronounced at short
distances, where the Lennard-Jones potential shows a much harder
behavior. This explains why Boek {\em et~al.} found a very large
dehydrated interlaminar space when they employed a Lennard-Jones
type pair potential for K-O \cite{Boek95a}. Naturally, they
overcome this difficulty by employing the pair potential proposed
by Bounds \cite{Bounds85}.

Nevertheless, since it is crucial for the hybrid Monte Carlo
simulations to keep the energy fluctuations as low as possible in
order to enlarge the acceptation rate \cite{Mehlig92}, it is
convenient to avoid employing relatively long range pair potential
contributions such as $\sim r^{-4}$, if no Ewald treatment is
applied on them. Hence, we refitted to equation \ref{Bounds} the
following expression
\begin{eqnarray} \label{fit}
U_{K-H_2O} \!\!&=& \!\! A_{KO}\exp{(-b_{KO}r_{KO})}-C_{KO}/r_{KO}^6 \nonumber \\
& & \!\! +A_{KH}\exp{(-b_{KH}r_{KH_1})} \nonumber \\ & & \!\!
+A_{KH}\exp{(-b_{KH}r_{KH_2})},
\end{eqnarray}
by employing a Levenberg-Marquardt algorithm and considering
several K-H$_{2}$O configurations. The procedure yields $A_{KO}=$
120750.2 kcal/mol, $b_{KO}=$ 3.4110 $\hbox{\AA}^{-1}$, $C_{KO}=$
5153.8 kcal$\hbox{\AA}^6$/mol, $A_{KH}=$ 2109.4 kcal/mol and
$b_{KH}=$ 2.8515 $\hbox{\AA}^{-1}$. We observed that avoiding the
$\sim r^{-4}$ term the acceptance rate enlarges more than three
times for a small (inner) time step of $0.8$ fs. In general,
both functions yield similar values of the K-H$_{2}$O potential
energy. Minima are located practically at the same distance
although the depth of the fitted function is 5$\%$ larger. For
larger distances this difference decreases. To check the
obtained interaction potential, a $NPT$ simulation containing 216
water molecules, a potassium cation, and a chloride anion was
performed at P
=1 atm and T=293 K. The corresponding K-O and K-H
radial distribution functions, g(r), and coordination numbers,
n(r), were studied. These functions were observed to be very
similar to those reported by Bounds \cite{Bounds85}. That is, the
K-O and K-H main peaks are located at 2.83 and 3.29 $\hbox{\AA}$
respectively, which compare well with their corresponding values
of 2.86 and 3.32 $\hbox{\AA}$ \cite{Bounds85}. The coordination
number for the g(r) minimum was found at 7.7 $\hbox{\AA}$, in
agreement with his value of 7.6 $\hbox{\AA}$ \cite{Bounds85}.
Moreover, all these values are even closer to the results obtained
by QM/MM simulations \cite{Tongraar98}. Hence, expression
\ref{fit} seems to be suitable for our purposes.

Finally, we should mention that electrostatic contributions, $\sim
r^{-1}$, were treated trough the implementation of the Ewald
summation formalism. Here the convergence factor was fixed to
$5.6/L_{min}$, where $L_{min}$ is the minimum simulation box side.
There were set five reciprocal lattice vectors for the directions
along the shortest sides and six vectors for the direction along
the largest side \cite{Alejandre94}. The dispersion-repulsion
contributions were corrected using the standard methods for
homogeneous fluids \cite{Allen} and a spherical cutoff of
$L_{min}/2$ was imposed.

\subsection{Simulations}
\label{simulation} Simulations were performed employing the hybrid
Monte Carlo (HMC) method \cite{Mehlig92,Odriozola_jcp}. This
technique allows making global moves while keeping a high average
acceptance probability. Global moves are done as follows from
molecular dynamics (MD), {\it i.~e.}, by assigning velocities and
by using a particular scheme for integrating the Newton's
equations of motion. Velocities are assigned randomly from a
Gaussian distribution in correspondence with the imposed
temperature, and in such a way that total momentum equals zero for
both interlaminar spaces. To fulfill detail balance condition, the
discretization scheme must be time reversible and area preserving
\cite{Mehlig92}. In particular, we employed the multiple time
scale algorithm given by Tuckerman and Berne \cite{Tuckerman92}.
This algorithm has the property of splitting the forces into short
and long range. The Lennard-Jones contribution plus the real part
of the electrostatic forces are set as short range and the
reciprocal space contribution of the electrostatic forces is set
as long range. To decrease time correlations a new configuration
is generated each 10 integration steps. The probability to accept
this new configuration is given by
\begin{equation}
P=\min\{1,\exp(-\beta \Delta \mathcal{H})\}
\end{equation}
where $\Delta \mathcal{H}$ is the difference between the new and
previous configuration Hamiltonians, and $\beta$ is the inverse of
the thermal energy. The long time step is set to 8 times the short
time step, and the short time step is chosen to obtain an average
acceptance probability of 0.7 \cite{Mehlig92}. This way, we
obtained short time steps close to 1.0 and 0.5 $fs$ for systems
containing 10 and 100 water molecules per interlaminar space,
respectively. As can be seen, the time step shortens with
increasing the system size, since energy fluctuations enlarge.
This is why HMC is not very efficient for systems counting on a
large number of movable sites. This is not our case, since few
ions and water molecules are the only contributors to energy
fluctuations. This makes HMC a reasonable choice. In fact, the
time steps we are obtaining are similar to those usually employed
for typical MD calculations
\cite{Skipper95c,Sposito01,Tambach04,Levesque02}.

For sampling in the $NP_{zz}T$ ensemble, after a trial change of
particles' positions, a box change is attempted in such a way that
the stress normal to the surface of the clays, $P_{zz}$, is kept
constant. For this purpose, box fluctuations are allowed only in
the $z$-direction and the probability for accepting the new box
configuration is given by
\begin{equation}
P\!=\! \min\{ 1,\exp [ - \beta ( \Delta \mathcal{U}+P_{zz} \Delta
V \! - N \beta ^{-1} \! \ln (V_n/V_o)) ] \}
\end{equation}
Here, $\Delta \mathcal{U}$ is the change in the potential energy,
$\Delta V$ is the volume change, $N$ is the total number of
molecules, and $V_n$ and $V_o$ are the new and old box volumes,
respectively \cite{dePablo01}.

For sampling in an open ensemble, the possibility of insertions
and deletions of water molecules has to be considered. Water
insertions and deletions were performed by Rosenbluth sampling
\cite{Hensen01,Odriozola_jcp}. The $\mu P_{zz}T$ ensemble
\cite{Odriozola_jcp} was used to obtain the equilibrium states
when the system is in contact with a reservoir at certain
temperature, pressure and water chemical potential. For this
purpose, the algorithm must sample the probability density of
finding the system in a particular configuration, {\it i.~e.}
\begin{equation}
\mathcal{N}_{\mu P_{z\!z}\!T} \! \propto \! \frac{ V^N \exp
\{-\beta [\mathcal{U}(\mathbf{s}^N)\!-\!\mu
N+\!P_{z\!z}V]\}}{\Lambda^{3N}N!}
\end{equation}
Hence, particle movements, insertions, deletions, and box changes
must be done as in a typical $NVT$, $\mu VT$ and $NP_{zz}T$
sampling \cite{Frenkel}. In particular, after trying a change of
particles' positions, we performed tries of inserting-deleting
water molecules. This is done by randomly calling both possible
trials in such a way that calls are equally probable. Since
accepting insertions or deletions are rare, we repeat this step 10
times or until any insertion or deletion is accepted. In case of
refusing the 10 insertion-deletion trials, we performed a box
trial move \cite{note}. In this way, the system rapidly evolves to
an equilibrium state. For some conditions, however, two free
energy local minima appear, which are accessed by handling initial
conditions.

\section{Results}
\label{results}

Results are presented in two subsections. These are {\it Sampling
in the $NP_{zz}T$ ensemble} and {\it Sampling in the $\mu P_{zz}T$
ensemble}. Each part presents the results for ground level
conditions, {\it i.~e.}, $T$$=$ 298 K and $P$$=$ 1 atm; and for 4
km of burial depth, {\it i.~e.}, $T$$=$ 394 K and $P$$=$ 600 atm,
assuming average gradients of 30 K/km and 150 atm/km.

\subsection{Sampling in the $NP_{zz}T$ ensemble}

\begin{figure}
\resizebox{0.45\textwidth}{!}{\includegraphics{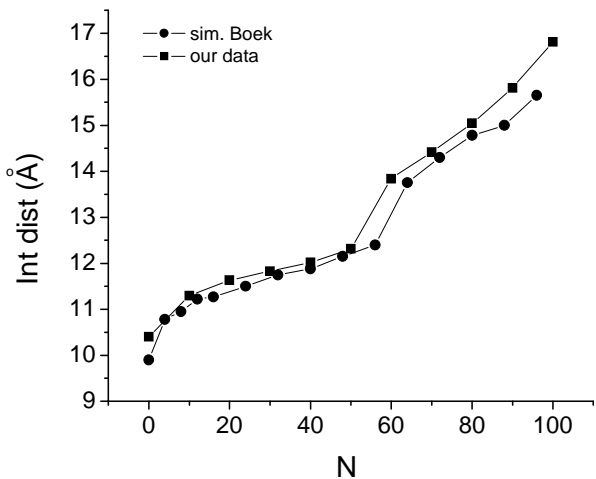}}
\caption{\label{NPT_swell_sup} Interlaminar distance as a function
of the number of water molecules per interlaminar space. }
\end{figure}

Let us first focus on the swelling behavior of the
K-montmorillonite hydrates under ground level conditions. This is
shown as a plot of the interlaminar distance as a function of the
number of water molecules obtained by $NP_{zz}T$ simulations. This
curve starts from a dehydrated state having a interlaminar
distance of 10.4 $\hbox{\AA}$, which is somewhat larger than the
experimental value of 10.1 $\hbox{\AA}$ \cite{Calvet73}. For a
small number of water molecules, the system reach interlaminar
distances above 11 $\hbox{\AA}$, which slightly increase with
increasing the number of water molecules producing a plateau. This
plateau shows interlaminar distances in the range 11.3 -- 12.3
$\hbox{\AA}$, which are clearly lower than those produced by
Na-montmorillonite \cite{dePablo01,Odriozola_lan}. This might be
seen as something unexpected, since potassium ions are much larger
than sodium ions. As we comment ahead, this is due to the different
interlayer structures that sodium and potassium ions generate. For
60 water molecules per interlaminar space, there is a jump from a
single water layer to a double water layer, which yields an
interlaminar distance of 13.9 $\hbox{\AA}$. For larger amounts of
water the system increases almost linearly, producing a very small
step when jumping from a double to a triple water layer structure.

Our findings are similar to those reported by Boek {\em et~al.}
\cite{Boek95a}. We included them in figure \ref{NPT_swell_sup} to
make easy the comparison. It is seen that trends are practically
equal. This indicates that differences in models and methods are
not very important. Nevertheless, they always obtain slightly
smaller interlaminar distances for a given number of water
molecules. This difference is always lower than 2$\%$, except for
the dehydrated state and the points close to 100 water molecules,
where differences close to 5$\%$ are observed. Since in our case
the jump from a single to a double layer is obtained for a lower
amount of interlaminar water, the points that correspond to 55-60
water molecules also show a larger difference. Anyway, differences
seem reasonable taking into account the differences in pair
potential definitions, box sizes, and methodologies.

\begin{figure}
\resizebox{0.28\textwidth}{!}{\includegraphics{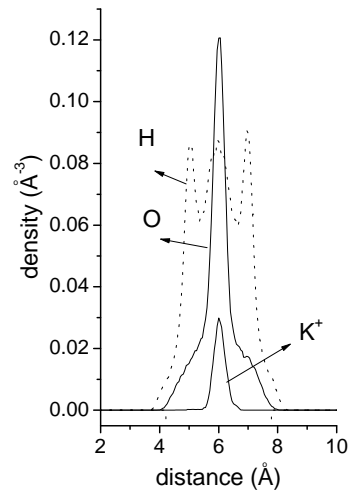}}
\caption{\label{perfil_ag40} Oxygen, hydrogen and potassium
density profiles of the interlaminar space. The water amount was
fixed to 40 molecules per interlaminar space. }
\end{figure}

\begin{figure}
\resizebox{0.45\textwidth}{!}{\includegraphics{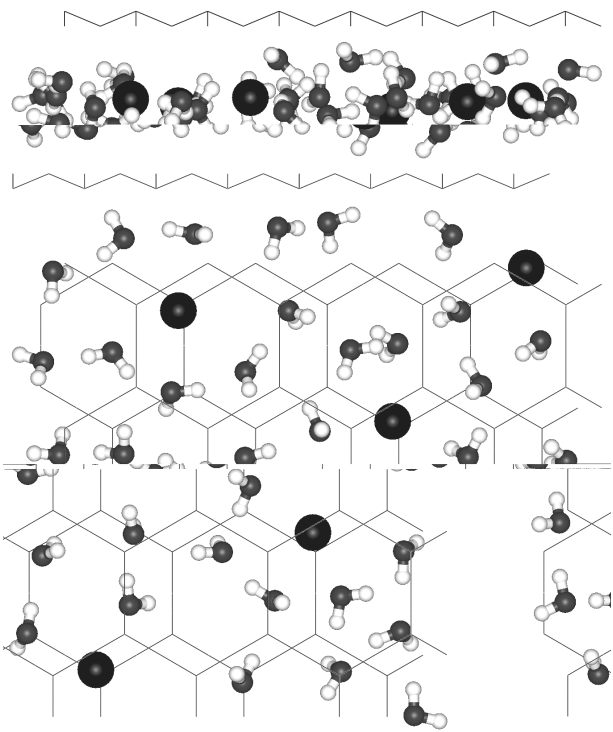}}
\caption{\label{lam_40} Snapshot of an equilibrated system having
40 water molecules. H sites are white, O are grey and K are black.
The wireframe represents the clay structure. The topmost image is
a side view and the lower image is the corresponding top view. }
\end{figure}

The structure of the interlaminar space for the system containing
40 water molecules per interlaminar space is shown in figure
\ref{perfil_ag40}. A high and narrow oxygen peak at the
interlaminar midplane, three hydrogen peaks, one coinciding with
the oxygen peak and the other two symmetrically situated at both
sides, and a single potassium ion peak, also at the interlaminar
midplane, are observed. Most of these results agree with others
previously reported \cite{Boek95a,Skipper98,Hensen01}. It should
be noted that this structure contrasts with that one of the
Na-montmorillonite single layer hydrate. In this case, sodium ions
are distributed at the sides of the oxygen peaks, closer to the
clay sheets, some of them being strongly attached to the clay
surface and so, forming inner-sphere surface complexes
\cite{Chang99}. This makes water molecules to cluster in two
layers joined at the interlaminar midplane, producing an oxygen
double peak \cite{dePablo04,Odriozola_jcp}. Naturally, this effect
widens the interlaminar space, despite of the smaller size of
sodium ion.

The radial distribution functions and coordination numbers for K-O
sites were also studied. Main g(r) peaks are situated at 2.77,
2.90, and 2.83 $\hbox{\AA}$ for water, clay, and total oxygen
sites, respectively. Although they are close to the one observed
for bulk potassium water solution, a contraction is seen for the
K-O water distance as a consequence of confinement. On the other
hand, the relatively large K-O clay distance is due to the
distribution of the potassium ions around the midplane of the
interlayer space. The coordination numbers for the first shell of
oxygen atoms are 5.1, 5.0, and 10.1 for water, clay, and total
oxygen sites, respectively. The clay and total coordination
numbers are somewhat inflated due to the smaller separation of the
O-O sites of the clay. This explains why the total coordination
number is larger than the one found for bulk. It should be noted
the large contribution of the clay to the total coordination
number. This suggests that potassium ions are interacting with
both clay sheets at the same time. In addition, the large K-O clay
distance may be indicating that potassium ions are directly
contributing to attract both layers toward the midplane, and
hence holding the sheets close to one another.

\begin{figure}
\resizebox{0.45\textwidth}{!}{\includegraphics{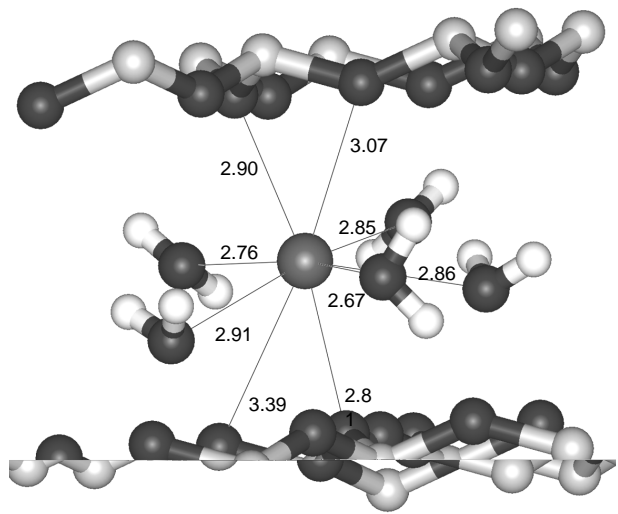}}
\caption{\label{coord-k} Zoom in of a potassium ion and its
coordination shell taken from figure \ref{lam_40}. Only water
molecules having K-O distances smaller than 3.7 $\hbox{\AA}$ are
shown. Distances in the figure are given in $\hbox{\AA}$. }
\end{figure}

All the already pointed features are illustrated in figure
\ref{lam_40}. For instance, it is observed that potassium ions are
practically centered on the interlayer midplane with an
approximate average of 5 water molecules surrounding each ion. It
is also possible to see how ions tend to separate from each other, and
that one of them is always close to a tetrahedral aluminum (these
sites are not highlighted in the figure). Moreover, it is shown that there are no
water molecules interposed between the potassium ions and the clay
sheets. Hence, potassium ions are behaving as inner-sphere
complexes, but simultaneously with both clay layers. To
see this even clearer, figure \ref{coord-k} was built by rotating
and zooming in figure \ref{lam_40}. Here, a potassium ion and its
inner water shell are shown. As can be seen, only 5 water
molecules surround the ion that coordinates with two oxygen atoms
from each clay. For this particular case, are also seen average
K-O distances of 2.81 and 3.04 $\hbox{\AA}$ for water and clay,
respectively.

\begin{figure}
\resizebox{0.28\textwidth}{!}{\includegraphics{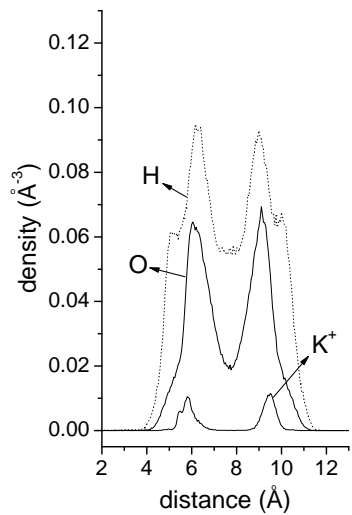}}
\caption{\label{perfil_ag80} Oxygen, hydrogen and potassium
density profiles of the interlaminar space. The water amount was
fixed to 80 molecules per interlaminar space. }
\end{figure}

As mentioned before, for 60 water molecules a double layer hydrate
is formed. This double layer hydrate becomes fully developed for
80 water molecules, where an interlayer distance close to 15.0
$\hbox{\AA}$ is observed. For this system, the oxygen, hydrogen,
and potassium profiles are shown in figure \ref{perfil_ag80}.
Here, the two oxygen peaks are a signature of the double water
layer structure. For each oxygen peak a potassium peak and two
hydrogen peaks are found. The potassium peak is found almost
coinciding with the oxygen peak, but slightly displaced at the
side closer to the clay sheet. This suggests the formation of
inner-sphere complexes. On the other hand, a small hydrogen peak
is found at the side closer to the clay sheet, and a larger one
closer to the interlaminar midplane. Hence, the third hydrogen
peak found for the one layer case is missing here.

\begin{figure}
\resizebox{0.45\textwidth}{!}{\includegraphics{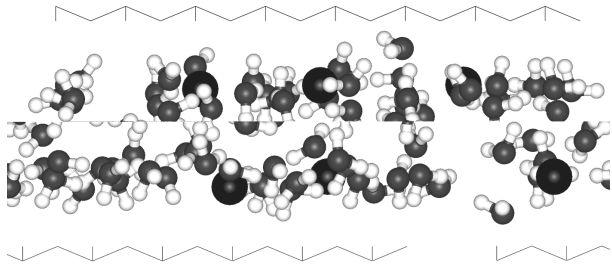}}
\caption{\label{lam_80} Snapshot of an equilibrated system having
80 water molecules per interlaminar space. H sites are white, O
are grey and K are black. The wireframe represents the clay
structure. }
\end{figure}

For this configuration, main g(r) peaks are located at 2.83, 2.75,
and 2.79 $\hbox{\AA}$ for water, clay, and total oxygen sites,
respectively. Hence, the K-O distance for water is not contracted
anymore, but equal to the bulk water K-O distance. On the
contrary, the K-O distance for the clay decreased 0.17
$\hbox{\AA}$. Moreover, this distance is found to be almost
constant for systems having more than 60 water molecules, strongly
suggesting that this is the natural average K-O distance for a
potassium ion attached to the siloxane surface. Hence, the
distance found for the single layer hydrate would be elongated as
a result of the pressure the water molecules produce on the clay
surfaces. The corresponding coordination numbers are 6.8, 2.8, and
9.6 for water, clay, and total oxygen sites, respectively. As can
be seen, the coordination number for the clay is much lower than
the value found for the single layer hydrate. This means that
potassium ions are coordinated to only one clay layer for the
double layer hydrate. Finally, and as expected, K-O water
coordination numbers increase with the number of water molecules,
whereas K-O clay and K-O total coordination numbers decrease. A
snapshot of this two layer hydrate is shown in figure
\ref{lam_80}. We should mention that the structure of the
inner-sphere complexes we found in this case are very similar to
those already reported by Sposito {\em et~al.} \cite{Sposito99}
(not shown here).

\begin{figure}
\resizebox{0.45\textwidth}{!}{\includegraphics{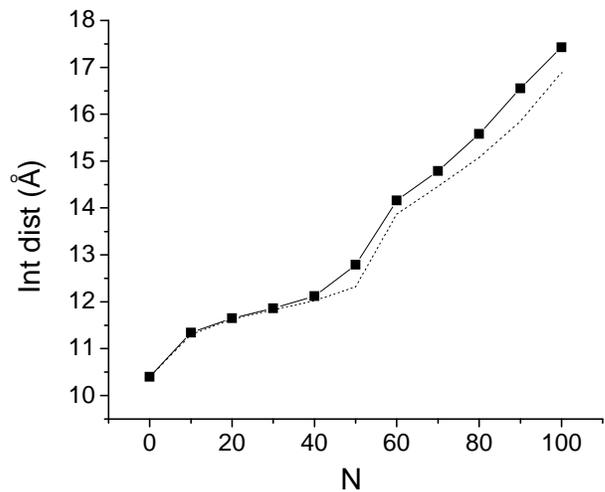}}
\caption{\label{NPT_swell_poz} Interlaminar distance as a function
of the number of water molecules per interlaminar space for burial
conditions. For comparison, the dotted line corresponds to ground
conditions (from figure \ref{NPT_swell_sup}). }
\end{figure}

\begin{figure}
\resizebox{0.28\textwidth}{!}{\includegraphics{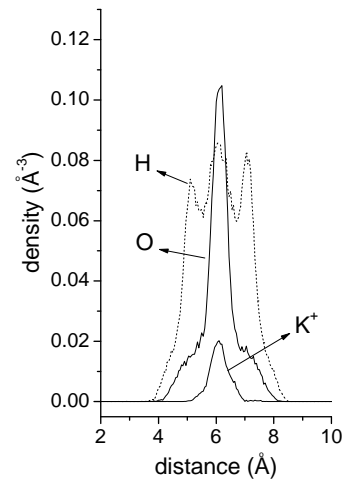}}
\caption{\label{perfil_ag40_poz} Oxygen, hydrogen, and potassium
density profiles of the interlaminar space. The water amount was
fixed to 40 molecules per interlaminar space for burial
conditions. }
\end{figure}

For burial conditions the interlaminar distance increases for a
given number of water molecules per interlaminar space. This is
shown in figure \ref{NPT_swell_poz}, where it shows the
interlaminar distance as a function of the number of water
molecules for burial and for ground level conditions. This
increment is more pronounced for large amounts of water. This is
an expected behavior since water molecules occupy larger effective
volumes at burial conditions \cite{Odriozola_lan}. Consequently,
profiles turn less sharp, {\it i.~e.}, peaks broad and shorten, as
it is shown in figure \ref{perfil_ag40_poz}. This was seen
experimentally by Skipper {\em et~al.}, although for a
Na-montmorillonite system \cite{Skipper00}. On the other hand,
trends for both the ground and burial data are very similar, {\it
i.~e.}, jumps are found at equal numbers of water molecules. This
last finding differs from the one found for Na-montmorillonite,
where the single layer to double layer jump occurs for 60 or 50
water molecules, depending on the burial depth
\cite{Odriozola_lan}.

\subsection{Sampling in the $\mu P_{zz}T$ ensemble}

Sampling in this ensemble allows the system to reach equilibrium
with a reservoir whose temperature, pressure, and, in our case,
water chemical potential are fixed. We employed the expression
$\beta \mu$ $=$ $\beta \mu_0 + \ln(p/p_0)$, where $p_0$ is the
vapor pressure at equilibrium with liquid water whose chemical
potential is $\mu_0$, and $p$ is the vapor pressure. For the TIP4P
water model with $T$=298 K and $P$=1 atm, we employed $\beta
\mu_0$=-17.4. This value was obtained by $NPT$ simulations of bulk
TIP4P water \cite{Odriozola_jcp} and using the Rosenbluth sampling
method explained elsewhere \cite{Hensen01,Frenkel}. It is in good
agreement with the value reported by Ch\'{a}vez-P\'{a}ez {\em
et~al.} \cite{dePablo01}, and it is 2.5$\%$ larger than the one
reported by Tambach {\em et~al.} \cite{Tambach04}. For $T$=394 K
and $P$=600 atm, $\beta \mu_0$=-13.4 was obtained
\cite{Odriozola_jcp}.

\begin{figure}
\resizebox{0.45\textwidth}{!}{\includegraphics{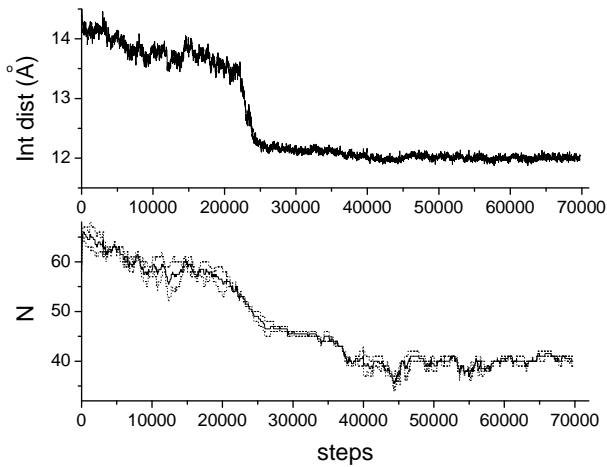}}
\caption{\label{int_n_step} Step evolution of the interlaminar
distance and number of water molecules obtained by $\mu P_{zz}T$
sampling. In the lower plot, dashed lines correspond to each
interlaminar space whereas the solid line is the average. Initial
conditions were 16 $\hbox{\AA}$ and 60 water molecules per
interlaminar space. Established conditions were T=298 K, P=1 atm,
and $p/p_o$=0.4. }
\end{figure}

The evolution of the interlaminar distance and number of water
molecules for ground level conditions and for $p/p_o$=0.4 is shown
in figure \ref{int_n_step}. Here, the initial conditions were 16
$\hbox{\AA}$ for the interlaminar distance and 60 water molecules
per interlaminar space. It is observed that the system reaches
approximately 65 water molecules and 14.3 $\hbox{\AA}$ of
interlaminar distance at the initial simulation steps. This
happens so fast since the system is initially very far from
equilibrium. Immediately after, the system starts slowly loosing
water molecules and decreasing its interlaminar distance on its
way toward equilibrium. It is observed that, once the system
reaches 13.5 $\hbox{\AA}$ of interlaminar space, it quickly falls
down to 12.2 $\hbox{\AA}$, loosing many water molecules during the
process. This is a signature of the transition from a double layer
hydrate to a single layer hydrate. The amount of water that
signals the transition goes from 55 to 45 molecules. This agrees
with the $NP_{zz}T$ results, where a single to a double layer
transition is observed in the range of 40 - 60 water molecules and
12.2 - 14.3 $\hbox{\AA}$. After the transition, it is observed
that the system takes several steps to finally reach equilibrium
at approximately 40000 steps. In this case, sampling was performed
in the step range of 40000 - 70000. For this particular run 12.0
$\hbox{\AA}$ of interlaminar space and 39.3 water molecules were
obtained.

\begin{figure}
\resizebox{0.45\textwidth}{!}{\includegraphics{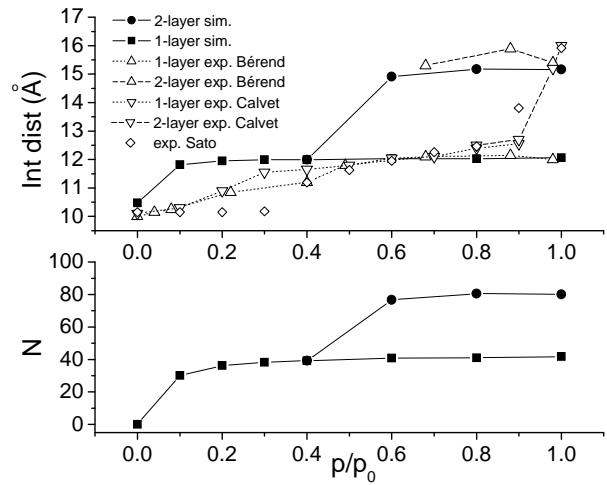}}
\caption{\label{MuPT_swell_sup} Interlaminar distance and number
of water molecules per interlaminar space as a function of the
vapor pressure for ground level conditions. Symbols {\tiny $\Box$}
and $\circ$ corresponds to initial conditions of 10 water
molecules and 11.5 $\hbox{\AA}$ of interlaminar distance, and 60
water molecules and 16.0 $\hbox{\AA}$ of interlaminar distance,
respectively. }
\end{figure}

Many runs were performed in order to build up figure
\ref{MuPT_swell_sup}. Here it is seen the interlaminar distance
and number of water molecules at equilibrium with reservoirs
having different water vapor pressures and for ground level
conditions. For zero vapor pressure, no matter what the
established initial conditions, the system is forced to eliminate
all of its water content, and so, the dehydrated state is yielded. This
has 10.40 $\hbox{\AA}$ of interlaminar space, as was found in the
preceding section. For increasing the vapor pressure, the system
starts to uptake water molecules from the reservoir producing
larger interlaminar distances. Nevertheless, the first water layer
saturates for vapor pressures over 0.2$p_0$ and so, a plateau is
generated. Again, this single layer hydrate is produced for
$p/p_0$ $\leq$ $0.4$ for all initial conditions.

For higher vapor pressures, however, at least two equilibrium
states are yielded. That is, depending on the established initial
conditions, the system may produce a double or a single layer
hydrate. Hence, the topmost lines of both plots of figure
\ref{MuPT_swell_sup} correspond to an initial configuration of 60
water molecules and the others to an initial configuration of 10
water molecules. These two equilibrium states are stable up to
water vapor saturation, producing an open hysteresis cycle. The
single layer state yields interlaminar distances ranging in 11.82
- 12.06 $\hbox{\AA}$ and amounts of water in the range of 30.1 -
41.8 molecules. The double layer state produces interlaminar
distances in the range 14.91 - 15.16 $\hbox{\AA}$ and numbers of
water molecules ranging in 76.7 - 80.6. These data contrast with
those obtained for Na-montmorillonite \cite{Odriozola_jcp}. In
this case, both initial conditions predict only a single layer
hydrate for $p/p_0$ $\leq$ $0.2$. In addition, this single layer
hydrate yields values of the interlaminar distance in the range of
12.34 - 12.73 $\hbox{\AA}$, having 34.6-47.5 water molecules
\cite{Odriozola_jcp}. Therefore, it becomes evident the
K-montmorillonite tendency to produce single layer hydrates even
for relatively high vapor pressures, and the small interlaminar
distance and amounts of water it yields. On the other hand, the
double layer plateaus for potassium and sodium do not behave very
differently for high vapor pressures.

We should also compare our results with the experimental data
obtained by B\'{e}rend {\em et~al.}\cite{Berend95},
Calvet\cite{Calvet73}, and Sato {\em et~al.}\cite{Sato92}. For
this purpose, these data are also included in figure
\ref{MuPT_swell_sup}. It should be pointed out that these data are
direct measurements of interlaminar distances against the relative
vapor pressure. This contrasts with the experimental data
presented in figure \ref{NPT_swell_sup}, which were obtained by
combining different experimental techniques and introducing some
assumptions. Here, B\'{e}rend {\em et~al.}'s data were obtained as
the evolution of the interlaminar distance for a dehydration
process, by starting from a double or a single layer hydrate.
Similarly, Calvet present his data by starting from a double or a
single layer hydrate, but for a hydration process. Sato {\em
et~al.}'s data are also obtained by hydration.

The best agreement between our simulations and experiments is
produced when comparing with the data of B\'{e}rend {\em et~al.}
\cite{Berend95} We highlight the double layer hydrate formation
for the range of relative vapor pressures of 0.6 - 1.0; the single
layer hydrate which keeps stable even for saturated vapor
pressures; the very good agreement for the basal space for the
single layer hydrate; and the relative good agreement for the
double layer hydrate interlaminar distance, close to 15.5
$\hbox{\AA}$. Calvet's data also are in general agreement with our
simulations. He observed a very stable single layer hydrate, with
interlaminar distances ranging in 11.8-12.5 $\hbox{\AA}$.
Nevertheless, he obtained a very unstable double layer hydrate.
Only for relative vapor pressures of 0.9 - 1.0 this hydrate was
observed. Its interlaminar distance is in the range of 15.2-16.0
$\hbox{\AA}$. Sato {\em et~al.}'s data also lead to similar
interlaminar distances for the single and double layer hydrate.
Finally, the single layer hydrate interlaminar distances also well
agree with data reported by Brindley and Brown \cite{Brindley}.
For 0.32, 0.52, and 0.79$p_0$ of vapor pressure they reported
11.9, 11.9 and 12.1 $\hbox{\AA}$, respectively.

On the other hand, the most important discrepancy with experiments
is that they obtained interlaminar distances close to 10.0
$\hbox{\AA}$ for the dehydrated state, which seems to be stable up
to a relative vapor pressure ranging from 0.1 to 0.3. We think that if
we were capable of reproducing the correct dehydrated distance, {\it
i.~e.}~10.0 instead of 10.4 $\hbox{\AA}$, this state would
probably become stable for small vapor pressures, as they found.

Experimentalists agree that for relative vapor pressures ranging
in 0.0 - 0.4, a mixture of the dehydrated state and the single
layer hydrate coexist, {\it i.~e.} there is interstratification.
Similarly, they observed the coexistence of double and single
layer hydrates, for high relative vapor pressures. In fact, Calvet
refers to his own data as \symbol{92}apparent distances";
B\'{e}rend {\em et~al.} conclude that all montmorillonites (they
studied Li, Na, K, Rb and Cs-montmorillonites) form
interstratified hydrates; and Sato {\em et~al.} observed several
\symbol{92}nonintegral basal reflections", which are interpreted
as \symbol{92}random or segregated-type interstratification of
collapsed and expanded layers". An example is the 13.81
$\hbox{\AA}$ of interlaminar distance he obtained for the relative
vapor pressure of 0.9. This point does not match either a single
or a double hydrate interlaminar distance, as it is clearly seen
in figure \ref{MuPT_swell_sup}. It is important to mention that
the general belief is that interstratification occurs due to
chemical heterogeneities of the clay layers. It was already proven
by simulations that changes on the positions of the clay
substitutions produce different interlaminar distances (although
differences are not very pronounced) \cite{dePablo01b}. Hence,
these heterogeneities surely lead to quasihomogeneous states, and
probably, to interstratified ones. On the other hand, a perfect
system like ours shows the double and the single layer hydrates to
be stable for identical conditions. Hence, we do not see any reason
for this not to occur in a real system. In other words, we think
that this is another source of interstratification, which arises
just as a consequence of the inherent thermodynamics of the
perfect system. In fact, similar conclusions are deduced by
Tambach {\em et~al.} \cite{Tambach04b}.

\begin{table}
\caption{\label{peaks} g(r) main K-O peak positions, $p$, and
first shell coordination numbers, $n$, for systems under different
water vapor pressures, $p/p_0$. Subindexes $w$ and $c$ refer to
water and clay, respectively. Peak positions are given in
$\hbox{\AA}$.}
\begin{ruledtabular}
\begin{tabular}{c|cccc|cccc}
& \multicolumn{4}{c|}{single water layer} & \multicolumn{4}{c}{double water layer} \\
$p/p_0$ & $p_w$ & $p_c$ & $n_w$ & $n_c$ & $p_w$ & $p_c$ & $n_w$ & $n_c$ \\
\hline 0.1 & 2.77 & 2.83 & 4.83 & 5.32 & -- & -- & -- & -- \\
0.2 & 2.77 & 2.85 & 4.96 & 5.11 & -- & -- & -- & -- \\
0.3 & 2.77 & 2.88 & 5.19 & 4.94 & -- & -- & -- & -- \\
0.4 & 2.77 & 2.90 & 5.17 & 4.96 & -- & -- & -- & -- \\
0.6 & 2.76 & 2.89 & 5.15 & 4.93 & 2.80 & 2.75 & 6.62 & 3.16 \\
0.8 & 2.76 & 2.88 & 5.24 & 4.85 & 2.82 & 2.77 & 6.77 & 2.82 \\
1.0 & 2.76 & 2.90 & 5.19 & 4.89 & 2.81 & 2.75 & 6.71 & 2.95 \\
\end{tabular}
\end{ruledtabular}
\end{table}

Table \ref{peaks} shows the main peak positions and the first
shell coordination numbers for the systems equilibrated at
different relative vapor pressures, and for a single and a double
water layer configurations. This was done to prove that the shifts
of the water and clay peaks mentioned in the preceding section are
indeed significant over a wide range of vapor pressures. As can be
seen for the single layer hydrate, the K-O water clay position is
practically constant and equal to the one found from the
$NP_{zz}T$ sampling. Also the double layer $p_w$ is close to the
value found in the previous section, {\it i.~e.}, similar to the
K-O bulk water position. On the other hand, $p_c$ increases for
increasing the relative water pressure, reaching a plateau for
$p/p_0$ $\geq$ 0.3. The plateau value is close to 2.89
$\hbox{\AA}$, which is also similar to the value found in the
preceding section. The decrease of this distance with $p/p_0$ is a
consequence of the decrease of the interlaminar distance with it.
For the double layer configuration it is also confirmed a value
close to 2.76 $\hbox{\AA}$. Hence, the shifts of these peaks are
relevant, and not just casual values obtained for some particular
conditions. From table \ref{peaks} it is seen that the values for
the coordination numbers also agree with those presented in the
previous section.

\begin{figure}
\resizebox{0.45\textwidth}{!}{\includegraphics{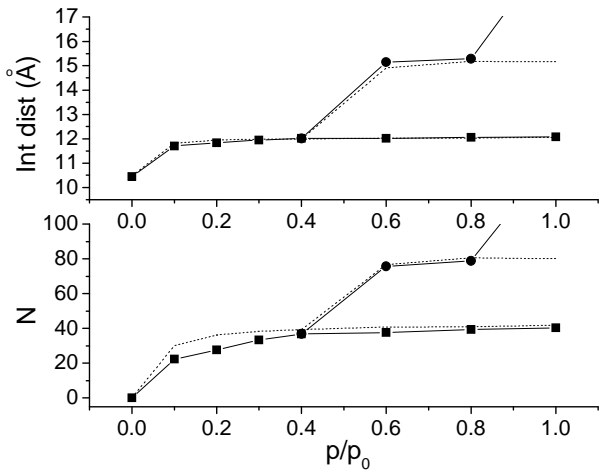}}
\caption{\label{MuPT_swell_poz} Interlaminar distance and number
of water molecules per interlaminar space as a function of the
vapor pressure for burial conditions. For comparison, the dotted
line corresponds to ground conditions (from figure
\ref{MuPT_swell_sup}). }
\end{figure}

To evaluate the effect of the burial depth on the
K-montmorillonite swelling curves, figure \ref{MuPT_swell_poz} was
built for $T=$394 K and $P=$600 atm. Additionally, results for
ground level conditions were included to make the comparison easy.
As can be seen, for initial conditions of 10 water molecules and
a small interlaminar space, the interlaminar distance is slightly
smaller than those obtained for ground level conditions. This
combines with the fact that water molecules occupy a larger
effective volume for burial conditions and so, the number of water
molecules decreases. In other words, the single layer hydrate of
K-montmorillonite dehydrates under burial conditions. Furthermore,
this single layer hydrate is stable even for a saturated water
vapor pressure. Again, this contrasts with the results we obtained
for Na-montmorillonite \cite{Odriozola_jcp}, where the single
layer hydrate was found to be unstable for large water activities.
Moreover, for $p/p_0=$ 1.0, the K-montmorillonite single layer
hydrate yields 12.07 $\hbox{\AA}$ of interlaminar space and 40.4
water molecules, which seems to be far from the transition range
of 45 to 55 water molecules previously found.

The double layer hydrate behaves differently. That is, it keeps
constant its water content and it slightly increases its
interlaminar distance. As found for ground level conditions, it
collapses forming the single layer hydrate for $p/p_0 <$ 0.6. On
the other hand, for a saturated vapor pressure the system
monotonically increases its amount of water as the simulation
evolves. In this last case we stopped the simulations when
reaching 180 water molecules, and thus, we assumed that the system
entered the osmotic regime.

In summary, these results indicate that once formed the
K-montmorillonite single water hydrate will not be destabilized
neither at ground level conditions nor at burial depths. In
addition, if the reservoir vapor pressure fells down 0.4$p_0$ the
single water hydrate will form.

\section{Discussion and Conclusions}
\label{summary}

Single and double layer hydrates of K-montmorillonite were studied
by means of $NP_{zz}T$ and $\mu P_{zz}T$ simulations. For that
purpose a hybrid Monte Carlo scheme was employed. Most results
from the $NP_{zz}T$ sampling just confirm those previously
reported elsewhere
\cite{Boek95a,Chang99,Hensen01,Sposito99,Park00}, suggesting that
differences in models and methods are not very important.

We found different clay-water-ion complexes for the single water
hydrate. They consist of an average of 5 water molecules
surrounding a potassium ion that additionally coordinates with two
oxygen atoms of each closest clay layer. These potassium ions are
placed at the interlayer midplane and so, distances between them
and the oxygens of the clay are slightly larger than those
observed for the double layer hydrate. It should be mentioned that
these midplane ions were reported by previous works as forming
outer-sphere complexes \cite{Boek95a,Skipper98}. In fact, Chang
{\em et~al.} \cite{Skipper98} found that these midplane ions have
a relatively large mobility, since potassium ions neither strongly
coordinate to water molecules nor strongly interact with the clay
surfaces. This is, indeed, characteristic of an outer-sphere
complex. This explains why they call them in this way.
Nonetheless, the inner-sphere definition given elsewhere
\cite{Chang99} says literally \symbol{92}The surface complex is
inner-sphere if the cation is bound directly to a cluster of
surface oxygen ions, with no water molecules interposed".
Accordingly, our midplane potassium ions form inner-sphere
complexes but simultaneously with both clay layers. Nevertheless,
we expect them to have a diffusivity similar to that reported by
Chang {\em et~al.} \cite{Chang99}. This is a clear difference
between this kind of inner-sphere complex and the typical
inner-sphere complex, where ions are strongly coordinated to one
of the surfaces \cite{Chang99}. This simultaneous coordination of
potassium ions with oxygen atoms of the two adjacent clay layers
plus their relatively large coordination distances, suggests to us
that these complexes are attracting the clay layers toward the
interlayer midplane, aiding to keep them together.

Although the previous finding seems to explain the stability of
the single layer hydrate found experimentally, we should also
reproduce computationally this behavior. Hence, $\mu P_{zz}T$
simulations were carried out. This ensemble lets interchange water
with a given reservoir and volume fluctuations. So, for any given
water vapor pressure of the reservoir, an average amount of water
and an average interlaminar distance are found. However, two
different equilibrium states may be produced, pointing to the
formation of free energy local minima \cite{Tambach04}, which are
accessed by handling initial conditions. This produces hysteresis
loops \cite{Tambach04}. We found that for initial conditions close
to the dehydrated state, a single water layer is always obtained
for any non zero vapor pressure, signaturing its high stability.
The amount of water of this single layer slightly increases as the
vapor pressure increases. This was obtained for ground level and
burial conditions. The interlaminar distance for the ground level
state is always close to 12.0 $\hbox{\AA}$. The number of water
molecules was found to be close to 38 for ground level conditions
and about 36 for burial conditions. This is due to the larger
effective volume the water molecules occupy at higher
temperatures.

On the other hand, the stability of the double layer hydrate
differs from that one of the single layer. For both conditions
studied, it was seen that the double layer collapses to form the
single layer for vapor pressures under 0.4$p_0$. In addition, this
double layer was found to be unstable for burial depth and for
saturated vapor pressures, producing a hydrated state in the
osmotic regime.

We should point out that these results agree with those of Boek
{\em et~al.} \cite{Boek95a}. In their figure 3 is shown the
potential energy of the interlayer water as a function of the
number of water molecules. It can be seen that the potassium curve
is very different than the sodium and lithium ones only for small
amounts of water (it shows much higher energy values, even well
above the water bulk reference). This supports our finding of an
extremely stable single water hydrate. Nevertheless, they claim
that potassium is a good swelling inhibitor due to its ability
to migrate and bind to the clay surfaces for large amounts of
water. Hence, the negatively charged surface becomes screened,
making its inherent repulsion less effective. We agree that this
mechanism explains the relatively high stability of double layer
hydrates, and even explains the stability of K-montmorillonite
hydrates in the osmotic regime. Nevertheless, it cannot explain
the remarkable stability of the single layer hydrate, since at
least a double layer is needed to obtain the binding between ions
and surfaces. This fact makes us think that midplane potassium
ions are playing an important role in the stability of single
layer hydrates.

Despite of thinking that the potassium simultaneously binding to
adjacent layers is the key to the single layer hydrate stability, we
do not think it is the only factor. Lowest interaction energy for
the pair K-TIP4P water is close to -20 kcal, which is higher than
that for Na-TIP4P of approximately -25 kcal. This means that
potassium ion does not strongly interact with water and therefore,
it easily loses water from its coordination shell. In other words,
the water chemical potential of the interlayer is not very
negative. That is, water may be more comfortable in bulk than in
the interlayer, surrounding the ions. As mentioned by Sposito {\em
et~al.} \cite{Sposito99}, potassium ions show a kind of
\symbol{92}hydrophobic character", in the sense that they tend to
interact with water molecules not only through their positive
charge but also through solvent cage formation. Consequently, some
of the water oxygen atoms may be easily exchanged by oxygen atoms
from the clay surface.

Finally, for drilling purposes and based on our results, it seems to
not be enough to add potassium to the mud in order to guarantee the
single layer water formation and, in this way, avoid swelling. One
should simultaneously decrease the mud water activity to safely
produce the single layer state. Once obtained, the clay will not
swell at all.

\section{Acknowledgments}
This research was supported by Instituto Mexicano del Petr\'{o}leo
grant D.00072.


\begin{thebibliography}{10}
\bibitem{note}
Different types of movements should be called randomly,otherwise detail balance
  is violated \cite{Frenkel}. The datapublished here and in
  references\cite{Odriozola_lan,Odriozola_jcp} were obtained by following
  thissequential algorithm. However, some data were checked by means ofthe
  corresponding random algorithm and the same results wereobtained.
\end{thebibliography}
\end{document}